\documentclass[
    ,final            
  ]
  {aipproc}

\layoutstyle{6x9}
\newcommand{\beq}{\begin{equation}}
\newcommand{\eeq}{\end{equation}}

\begin{document}

\title{MSSM inflaton:  SUSY dark matter and LHC}


\author{A. Mazumdar}{
  address={Physics Department, Lancaster University, Lancaster, LA1 4YB, UK\\
  Niels Bohr Institute, Copenhagen University, Blegdamsvej-17, DK-2100, Denmark }
}

\begin{abstract}
In this talk we will discuss how inflation can be embedded
within a minimal extension of the Standard Model where the inflaton
carries the Standard Model charges. There is no need of an ad-hoc
scalar field to be introduced in order to explain the temperature
anisotropy of the cosmic microwave background radiation, all the
ingredients are present within a minimal supersymmetric Standard
Model. For the first time inflaton properties can be directly linked to
the particle phenomenology, dark matter, and the baryons of the Standard Model.
\end{abstract}

\maketitle

Inflation has been extremely successful in explaining the temperature
anisotropy of the observed comsic microwave background radiation by
generating almost scale invariant density perturbations~\cite{WMAP3}.
It has been known that inflation can be driven by a
dynamical scalar field known as the {\it inflaton}, an order
parameter, which could either be fundamental or
composite. Particularly, if the inflaton rolls very slowly on a
sufficiently flat potential such that the potential energy density
dominates over the kinetic term, then all the successes of inflation
can be met, i.e. dynamical explanation of the homogeneity and the
isotropy of the universe on very large scales, nearly Gaussian density 
perturbations, etc.

Inspite of the impressive list of achievements, it has been proven hard to embed 
inflation in a fundamental theory which could also be testable in
a laboratory. In past one has always relied on scalar fields
which are {\it absolute gauge singlets} possibly residing in a
hidden sector or a secluded sector with an unknown  couplings to the SM gauge
group. By definition, an {\it absolute gauge singlet} does not carry
any charge what so-ever be the case, therefore, the masses, couplings
and interactions are not generally tied to any fundamental theory or
any symmetry.  Such gauge singlets are used ubiquitously by model
builders to obtain a desired potential and interactions {\it almost} at a free will
in order to explain the current CMB data. 

Very recently some of these questions have been addressed in a low
energy field theory setup, which explains (for a review see
\cite{anupam}): a) the origin of inflation,~ b) the fundamental interactions 
of an inflaton,~c) how the inflaton creates Standard Model baryons and cold dark matter ?,
and d) how can we test the inflaton in a laboratory ?

For the first time we have been successful in constructing an 
inflaton which carries the Standard Model (SM) 
charges and embedded within supersymmetry(SUSY)~\cite{AEGM,AEGJM,AKM,AJM}. 
The SUSY provides the scalar fields (partners of the SM fermions and gauge
bosons) and the stability of the flatness of the inflaton potential.
Since the inflaton carries the SM charges, it 
decays {\it only} into the SM particles and SUSY particles, i.e. quarks, squarks,
leptons, sleptons, etc. Within a minimal supersymmetric Standard Model
(MSSM) we know all the relativistic species and therefore we can trace
back thermal history of the universe accurately above the electroweak
scale.  If the lightest
supersymmetric particle (LSP) is stable due to R-parity, we naturally
obtain cold dark matter after the MSSM inflation~\cite{ADM}.

\begin{figure}
 \includegraphics[height=.20\textheight]{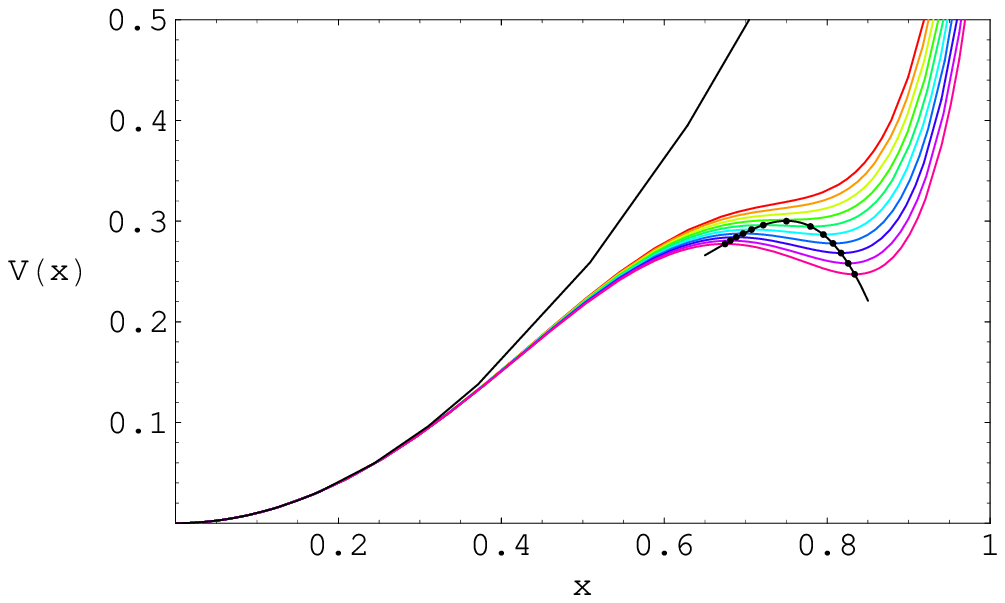}~~~~~~~~~~~
 \includegraphics[height=0.28\textheight]{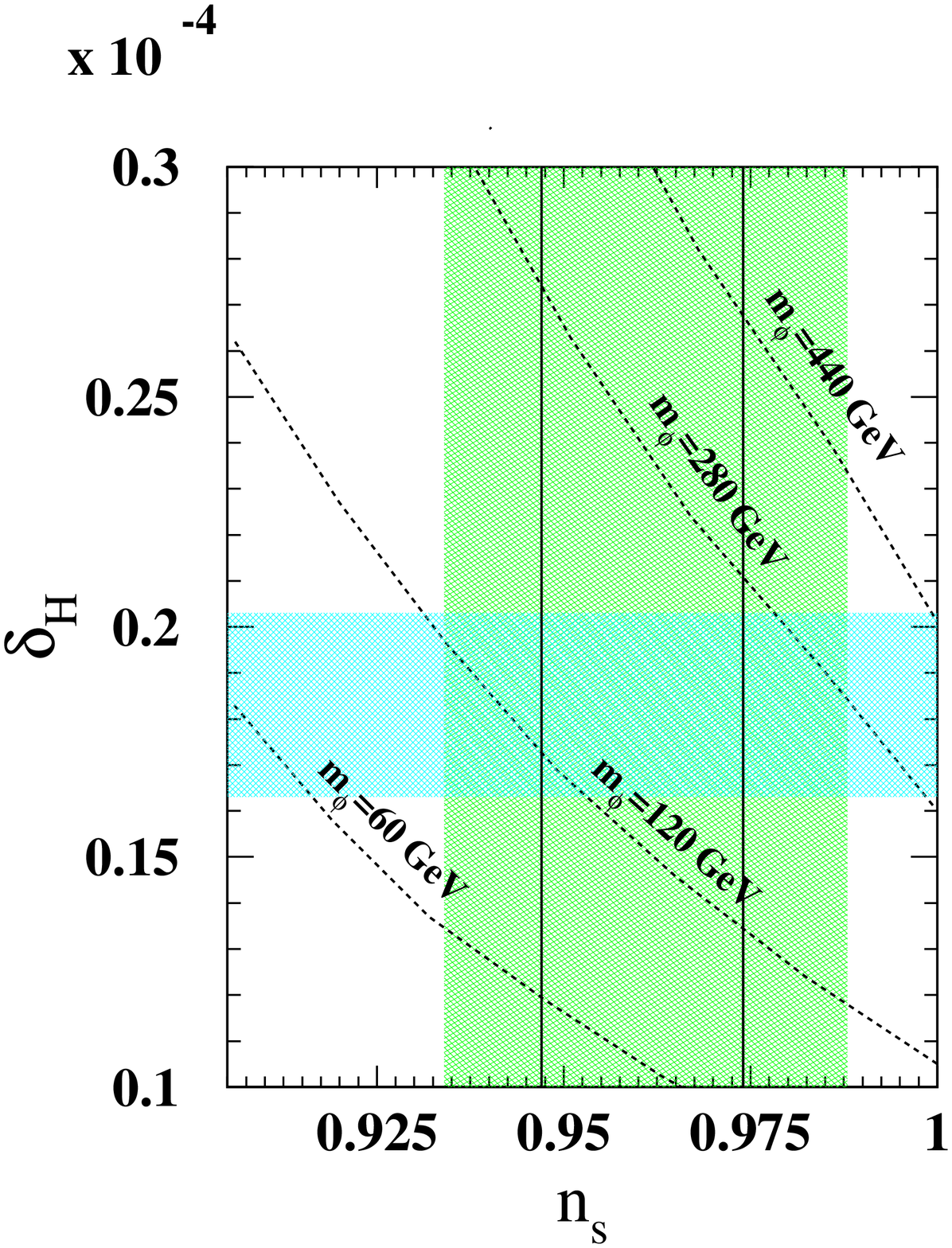}  
 \caption{The shape of the potential on the left hand side. On the right hand side, we have shown the 
 constraints on the mass of the inflaton, $m_{\phi}(\phi_0=10^{14}~{\rm GeV})$,
 from the amplitude, $\delta_{H}$, and tilt, $n_s$, of the power spectrum. Note that the 
 inflaton is made up of either $udd$ or $LLe$, see Ref.~\cite{ADM3}.}\label{nsdel0}
 \end{figure}

Let us now mention the main features of MSSM flat direction inflation~\cite{AEGM,AEGJM}. There are {\it only}
two flat directions,  $udd$ and $LLe$, which are lifted by higher order superpotential terms of the following form
\cite{MSSM-REV,GKM}:
\beq \label{supot}
W \supset {\lambda \over 6}{\Phi^6 \over M^3_{\rm P}}\, ,
\eeq
where $\lambda \sim {\cal O}(1)$, rest of the flat directions within MSSM are lifted by hybrid superpotential
terms, i.e. $W\supset (\Phi^{n-1}\Psi)/M_{\rm P}^{n-3}$. Such terms yield no non-renormalizable A-term in the potential, which is relevant for 
our discussion below. The scalar component of $\Phi$ superfield,  denoted by $\phi$, is given by
\beq \label{infl}
\phi = {{u} + {d} + {d} \over \sqrt{3}} ~ ~ ~ , ~ ~ ~ \phi = {{L} + {L} + {e} \over \sqrt{3}},
\eeq
for the $udd$ and $LLe$ flat directions respectively.

Writing the complex scalar field $\Phi$ in terms of radial and angular components $\Phi = \phi {\rm exp}[i \theta]$, the scalar potential along the radial direction $\phi$ is found to be~\cite{AEGM,AEGJM}
\beq \label{scpot}
V(\phi) = {1\over2} m^2_\phi\, \phi^2 - A {\lambda\phi^6 \over 6\,M^{6}_{\rm P}} + \lambda^2
{{\phi}^{10} \over M^{6}_{\rm P}}\,,
\eeq
where $m_\phi$ and $A$ are the soft breaking mass and the $A$-term respectively. We have already minimized the potential along the angular direction $\theta$. Note that $A$ is a positive quantity as we can absorb its phase by a redefinition of 
$\theta$. Provided that
${A^2 \over 40 m^2_{\phi}} \equiv 1 + 4 \alpha^2$,
where $\alpha^2 \ll 1$, there exists a point of inflection in $V(\phi)$~\cite{ADM,AM}
\begin{eqnarray}
\phi_0 = \left({m_\phi M^{3}_{\rm P}\over \lambda \sqrt{10}}\right)^{1/4} + {\cal O}(\alpha^2) \,,~~~~~~~~~~~~~
V^{\prime \prime}(\phi_0) = 0 \, , \label{2nd}
\end{eqnarray}
at which
\begin{eqnarray}
\label{pot}
V(\phi_0) = \frac{4m_{\phi}^2\phi_0^2}{15} + {\cal O}(\alpha^2) \, ,
V'(\phi_0) = 4 \alpha^2 m^2_{\phi} \phi_0 \, + {\cal O}(\alpha^4) \,,
V^{\prime \prime \prime}(\phi_0) = \frac{32m_{\phi}^2}{\phi_0} + {\cal O}(\alpha^2) \, . 
\end{eqnarray}
From now on we only keep the leading order terms in all expressions. Note that in gravity mediated SUSY breaking, the $A$-term and the soft SUSY breaking mass are of the same order of magnitude as the gravitino mass, i.e. $m_{\phi} \sim A \sim m_{3/2} \sim (100~{\rm GeV}-1~{\rm TeV})$. Therefore the above conditions can indeed be satisfied. We then have 
$\phi_0 \sim {\cal O}(10^{14}~{\rm GeV})$. Inflation occurs within an interval,
$\vert \phi - \phi_0 \vert \sim {\phi^3_0 \over 60 M^2_{\rm P}} $,
in the vicinity of the point of inflection, within which the slow roll parameters $\epsilon \equiv (M^2_{\rm P}/2)(V^{\prime}/V)^2$ and $\eta \equiv M^2_{\rm P}(V^{\prime \prime}/V)$  are smaller than $1$. The Hubble expansion rate during inflation is given by

\beq \label{hubble}
H_{\rm MSSM} \simeq \frac{1}{\sqrt{45}}\frac{m_{\phi}\phi_0}{M_{\rm P}} 
\sim (100~{\rm MeV}-1~{\rm GeV})\,.
\eeq
The amplitude of density perturbations $\delta_H$ and the scalar spectral index $n_s$ are given by~\cite{AEGJM,ADM,ADM3,AM}:
\beq \label{ampl}
\delta_H = {8 \over \sqrt{5} \pi} {m_{\phi} M_{\rm P} \over \phi^2_0}{1 \over \Delta^2}
~ {\rm sin}^2 [{\cal N}_{\rm COBE}\sqrt{\Delta^2}]\,,~~~
n_s = 1 - 4 \sqrt{\Delta^2} ~ {\rm cot} [{\cal N}_{\rm COBE}\sqrt{\Delta^2}], \eeq
where
\beq \label{Delta}
\Delta^2 \equiv 900 \alpha^2 {\cal
N}^{-2}_{\rm COBE} \Big({M_{\rm P} \over \phi_0}\Big)^4\,. \eeq
${\cal N}_{\rm COBE}$ is the number of e-foldings between the time when the observationally 
relevant perturbations are generated till the end of inflation and follows: ${\cal
N}_{\rm COBE} \simeq 66.9 + (1/4) {\rm ln}({V(\phi_0)/ M^4_{\rm P}}) \sim 50$~\cite{MULTI}.
We note that reheating after MSSM inflation is very fast, due to gauge couplings of the inflaton to gauge/gaugino fields, and results in a radiation-dominated universe within few Hubble times after the end of inflation~\cite{AEGJM,RA}.



A remarkable property of MSSM inflation, which is due to inflation occurring near a point of inflection, is that it can give rise to a wide range of scalar spectral index. This is in clear distinction with other models (for example, chaotic inflation, hybrid inflation natural inflation, etc.) and makes the model very robust. Indeed it can yield a spectral index within the whole $2 \sigma$ allowed range by 5-year WMAP data $0.934 \leq n_s \leq 0.988$~\cite{WMAP3}. This happens for
\beq \label{Delta2}
0 \leq \Delta^2 \leq {\pi^2 \over 4 {\cal N}^2_{\rm COBE}}\,. \eeq
In Fig.~(\ref{nsdel0}), we show $\delta_H$ as a function of $n_s$ for
different values of $m_{\phi}(\phi_0)$. The horizontal blue band shows the $2 \sigma$ 
experimental band for $\delta_H$. The vertical green shaded region is the 2$\sigma$ experimental band for $n_s$. The region enclosed by  solid lines shows the 1$\sigma$ experimental allowed region.
We find that smaller values of $m_{\phi}$ are preferred for smaller values of $n_s$. We also find
that the allowed range of $m_{\phi}$ is $90-330$~GeV for the
experimental ranges of $n_s$ and $\delta_H$. This figure is drawn for $\lambda \simeq
1$, which is natural in the context of effective field theory (unless it is suppressed 
due to some symmetry). Smaller values of $\lambda$ will lead to an increase in $m_{\phi}$~\cite{ADM}.

Initial condition for inflation has been studied in Ref.~\cite{AFM} and in Ref.~\cite{ADM3}, we 
showed that MSSM inflaton is always driven to the point of inflection if there is an earlier 
false vacuum phase of inflation. In this respect, although inflation occurs at a low scale, the 
initial conditions are {\it natural} once the MSSM inflation is preceded by a high scale inflation, i.e.
$H_{inf}\geq H_{\rm MSSM}$.
 We also showed how to realize such a false vacuum inflation within MSSM.
Recently we have also shown that the MSSM inflaton will fragment after inflation, and the process of fragmentation will
leaves its imprints in the gravity waves, with a detectable frequency and amplitude which matches with that of  
LISA and LIGO-III experiments~\cite{KM}.

Since $m_{\phi}$ is related to the scalar masses, sleptons ($LLe$
direction) and squarks ($udd$ direction), the bound on $m_{\phi}$
will be translated into the bounds on these scalar masses which are
expressed in terms of the model parameters~\cite{AEGJM}. The models
of mSUGRA depend only on four parameters and one sign. These are
$m_0$ (the universal scalar soft breaking mass at the GUT scale
$M_{\rm G}$); $m_{1/2}$ (the universal gaugino soft breaking mass at
$M_{\rm G}$); $A_0$ (the universal trilinear soft breaking mass at
$M_{\rm G}$); $\tan\beta =\langle H_2 \rangle \langle H_1 \rangle$ at the electroweak scale
(where $H_2$ gives rise to $u$ quark masses and $H_1$ to $d$ quark
and lepton masses); and the sign of $\mu$, the Higgs mixing
parameter in the superpotential ($W_{\mu} = \mu H_1 H_2$).
Unification of gauge couplings within supersymmetry suggests that
$M_{\rm G} \simeq 2 \times 10^{16}$ GeV. The model parameters are
already significantly constrained by different experimental results.
Most important constraints are: The light Higgs mass bound of 
$M_{h^0} > 114.0$~GeV from LEP~\cite{higgs1}. The $b \rightarrow s \gamma$ 
branching ratio: $2.2\times10^{-4} < {\cal B}(B \rightarrow X_s \gamma) <
4.5\times10^{-4}$~\cite{bsgamma}. In mSUGRA the $\tilde\chi^0_1$ is the candidate for CDM.  The
$2\sigma$ bound from the WMAP gives a relic density bound
for CDM to be $0.095 < \Omega_{\rm CDM} h^2 < 0.129 $~\cite{WMAP3}.
The bound on the lightest chargino mass of
$M_{\tilde\chi^{\pm}_1} > 104$~GeV from LEP~\cite{aleph}. The possible $3.3~\sigma$ deviation
(using $e^+e^-$ data to calculate the leading order hadronic contribution)from the SM expectation of the
anomalous muon magnetic moment from the muon $g-2$ collaboration~\cite{BNL}.

The allowed mSUGRA parameter space, at present, has mostly three
distinct regions: (i)~the stau-neutralino
($\tilde\tau_1~-~\tilde\chi^1_0$), coannihilation region where
$\tilde\chi^1_0$ is the lightest SUSY particle (LSP), (ii)~the
$\tilde\chi^1_0$ having a dominant Higgsino component (focus point)
and (iii)~the scalar Higgs ($A^0$, $H^0$) annihilation funnel
(2$M_{\tilde\chi^1_0}\simeq M_{A^0,H^0}$)~\cite{darkrv}. These three regions have
been selected out by the CDM constraint. There stills exists a bulk
region where none of these above properties is observed, but this
region is now very small due to the existence of other experimental
bounds. After considering all these bounds we will show that there
exists an interesting overlap between the constraints from inflation
and the CDM abundance.

We calculate $m_{\phi}$ at $\phi_0$ and $\phi_0$ is $~10^{14}$ GeV
which is two orders of magnitude below the GUT scale. From this
$m_{\phi}$, we determine $m_0$ and $m_{1/2}$ by solving the RGEs for
fixed values of $A_0$ and $\tan\beta$.
After we determine $m_0$ and $m_{1/2}$ from $m_{\phi}$, we can
determine the allowed values of $m_{\phi}$ from the experimental
bounds on the mSUGRA parameters space. In order to obtain the
constraint on the mSUGRA parameter space, we calculate the SUSY
particle masses by solving the RGEs at the weak scale using four
parameters of the mSUGRA model and then use these masses to
calculate Higgs mass, $BR[b\rightarrow s \gamma]$, dark matter
content etc.

\begin{figure}
\includegraphics[height=.3\textheight]{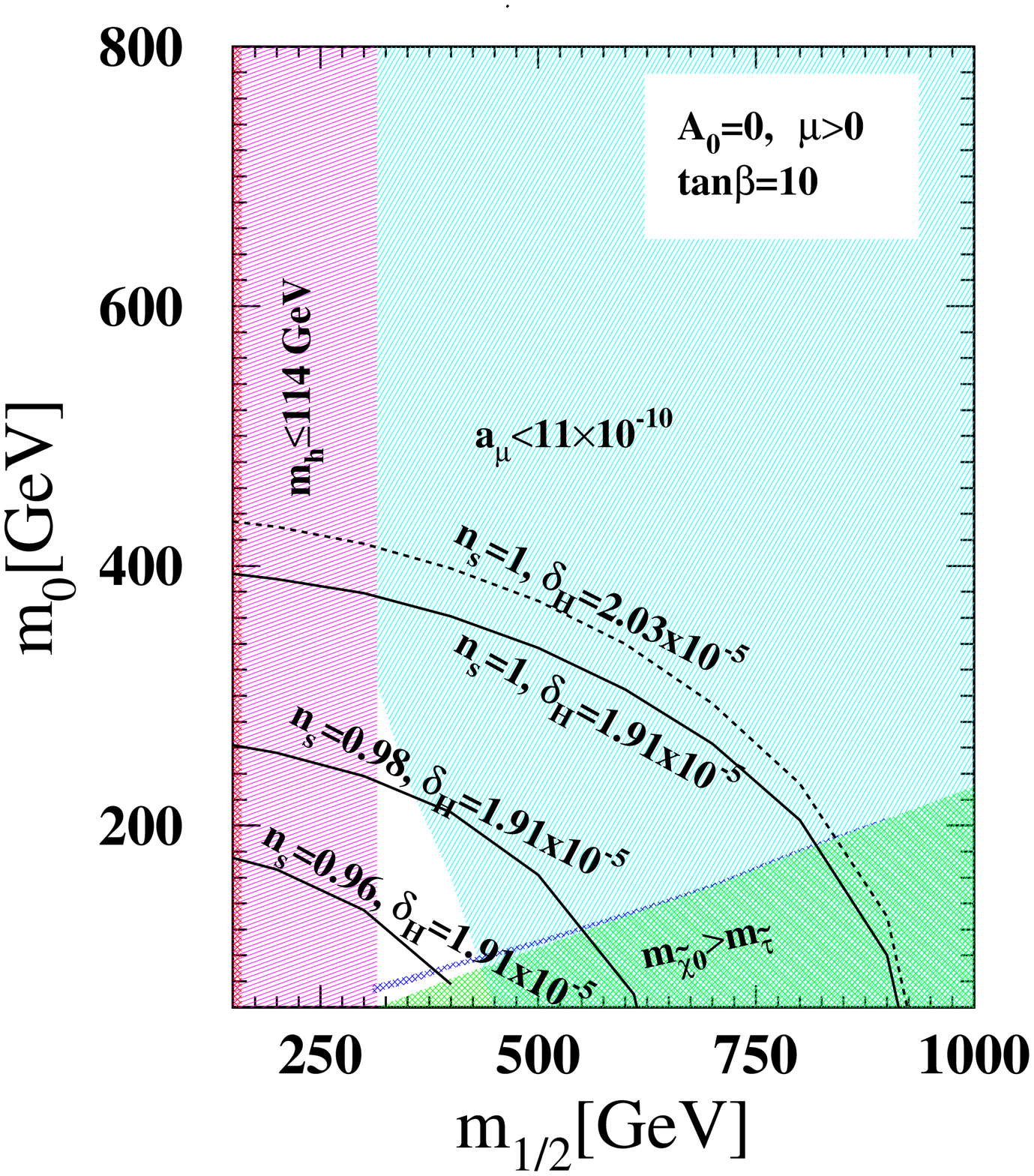}\,~~~~~~~~~~~
\includegraphics[height=.3\textheight]{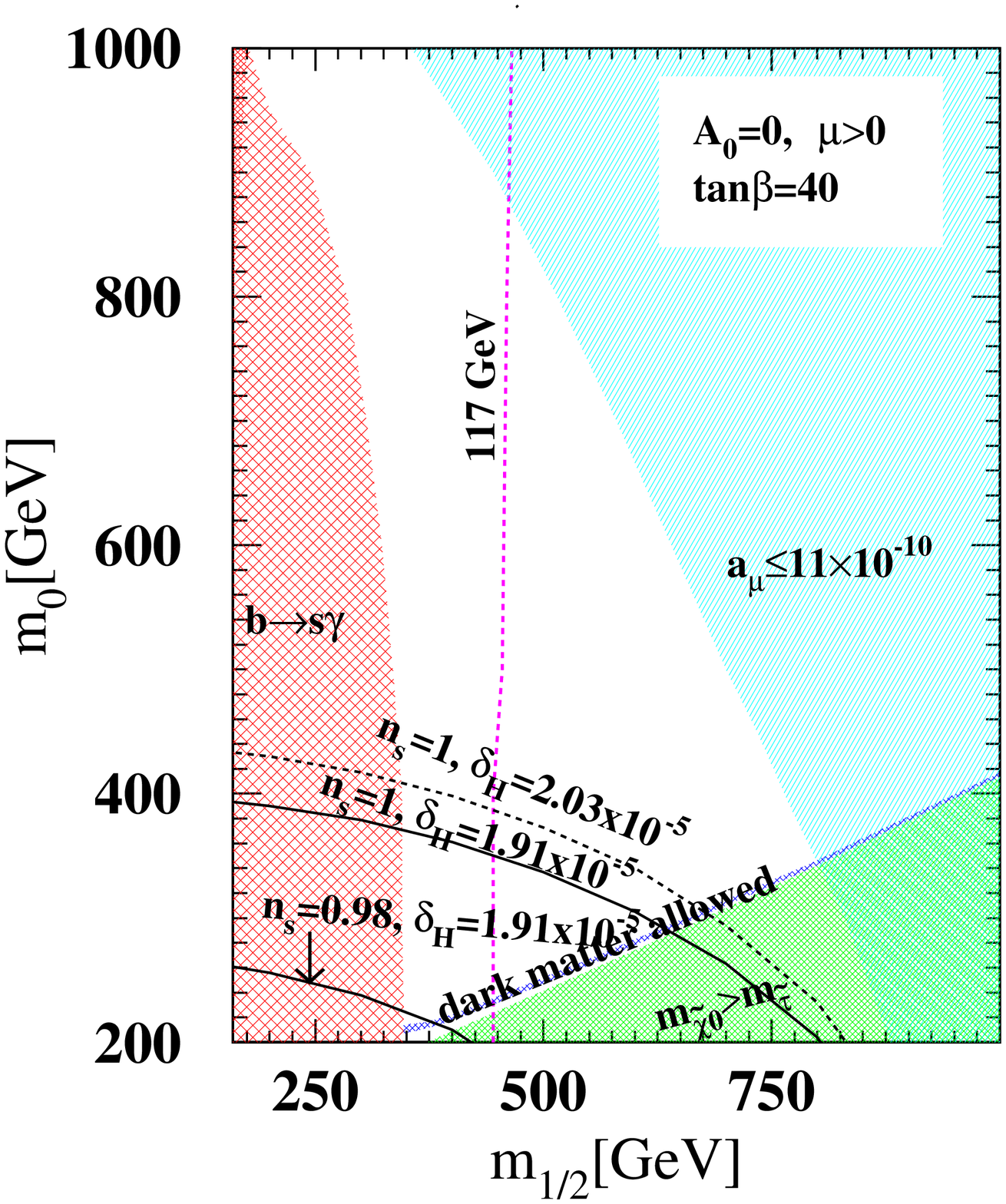} 
\caption{The
contours for different values of $n_s$ and $\delta_H$ are shown in
the $m_0-m_{1/2}$ plane for $\tan\beta=10$ and $tan\beta =40$. We 
used $\lambda=1$ for the contours. We show the dark matter allowed 
region {narrow blue corridor}, (g-2)$_\mu$ region (light blue) for $a_{\mu}\leq
11\times10^{-8}$, Higgs mass $\leq 114$ GeV (pink region) and LEPII
bounds on SUSY masses (red). We also show the the dark matter
detection rate by vertical blue lines. In the right hand panel, we 
$b\rightarrow s\gamma $ allowed region (brick).}\label{10flat}
\end{figure}

We show that the mSUGRA parameter space in Figs.~(\ref{10flat}) 
for $\tan\beta=10$ and $40$ with the $udd$ flat
direction using $\lambda=1$. In the
figures, we show contours correspond to $n_s=1$ for the maximum
value of $\delta_H=2.03\times 10^{-5}$ (at $2\sigma$ level) and
$n_s=1.0,~0.98,~0.96$ for $\delta_H=1.91\times 10^{-5}$. The
constraints on the parameter space arising from the inflation
appearing to be consistent with the constraints arising from the
dark matter content of the universe and other experimental results.
 We find that $\tan\beta$ needs to
be smaller to allow for smaller values of $n_s<1$. It is also
interesting to note that the allowed region of $m_{\phi}$, as
required by the inflation data for $\lambda=1$ lies in the
stau-neutralino coannihilation region which requires smaller values
of the SUSY particle masses. The SUSY particles in this parameter
space are, therefore, within the reach of the LHC very quickly. The
detection of the region at the LHC has been considered in
refs~\cite{dka}. From the figures, one can also find that as
$\tan\beta$ increases, the inflation data along with the dark
matter, rare decay and Higgs mass constraint  allow smaller ranges
of $m_{1/2}$. For example, the allowed ranges of  gluino masses are
765 GeV-2.1 TeV and 900 GeV-1.7 TeV for $\tan\beta=10$ and 40
respectively. 

There are other relevant cases which we have not discussed here, we refer
the readers to Ref.~\cite{ADM}. The dark matter abundance has also been studied when
the inflaton is a gauge invariant combination of $NH_{u}L$, first studied in Ref.~\cite{AKM},
and the dark mater analysis was done in Ref.~\cite{ADM2}.

To summarize our analysis provides an example of a Standard Model {\it gauge
invariant} inflaton giving rise to a successful inflation and explains
the neutralino CDM abundance, which is in agreement with the present
cosmological observations. Moreover this is the first example where
the ingredients of a primordial inflation can be put onto test in a
laboratory physics such as in the case of LHC.

\begin{theacknowledgments}
 A.M is thankful to the organizers for their kind invitation. The author is also thankful to R. Allahverdi, 
 B. Dutta, K. Enqvist, J. Garcia-Bellido and A. Jokinen and A. Kusenko for helpful discussions.
 
 \end{theacknowledgments}


\end{document}